\documentclass[12pt,epsbox]{article}
\usepackage{color}
\usepackage[dvips]{graphicx,psfrag}
\usepackage{amsmath,amssymb}
\usepackage{bm}
\baselineskip=\normalbaselineskip
\renewcommand{\baselinestretch}{1.3}
\renewcommand{\thefootnote}{\fnsymbol{footnote}}

\begin{document}
\setcounter{page}{0}
\begin{flushright}
\parbox{40mm}{%
KUNS-1890 \\
{\tt hep-th/0401153} \\
January 2004}

\end{flushright}

\vspace{1cm}

\begin{center}
{\large{\bf 
Noncommutative inflation and the large-scale\\ damping in the CMB anisotropy%
\footnote{
To appear in the Proceedings of the 3rd Symposium on 
Quantum Theory and Symmetries (QTS3), Cincinnati, Ohio, 
10--14 Sept 2004 --- \copyright~World Scientific.
}}}
\end{center}

\vspace{5mm}

\begin{center}
{\sc Masafumi Fukuma}\footnote%
{E-mail: {\tt fukuma@gauge.scphys.kyoto-u.ac.jp}},  
{\sc Yuji Kono}\footnote%
{E-mail: {\tt kono@gauge.scphys.kyoto-u.ac.jp}} and 
{\sc Akitsugu Miwa}\footnote%
{E-mail: {\tt akitsugu@gauge.scphys.kyoto-u.ac.jp}}  \\[0.6em]
{\sl Department of Physics, Kyoto University, Kyoto 606-8502, Japan} \\

\end{center}

\vspace{5mm}
\renewcommand{\thefootnote}{\alph{footnote}}
\setcounter{footnote}{0}
\addtocounter{page}{1}

\begin{center}
{\bf abstract}
\end{center}

\begin{quote}
We show that a certain class of short-distance cutoff 
can give rise to large suppression 
on the CMB anisotropies at large angular scales.
\end{quote}

\vspace{3mm}
\renewcommand{\baselinestretch}{1.4}


\section{\large{Introduction}}

The inflationary models of the universe 
give a desired initial condition to the standard 
big bang cosmology. 
They are supported by the recent elaborate measurement 
of the CMB anisotropy \cite{data},  
except that the observed angular power spectrum of the CMB 
for small multipoles $l$ ($l\lesssim 10$) 
have much smaller values than the theoretical prediction. 
The standard explanation of this discrepancy at large angular scales 
is based on the cosmic variance \cite{Liddle}.  
However, if this is not simply a statistical deviation, 
then there is a possibility that this is a remnant of string dynamics 
in the very early universe.

One of the basic results in string theory 
is the existence of the minimum length scale $l_s$ 
\cite{minimum_length}, 
and spacetime is expected to loose its smooth Riemannian structure 
and to become discrete (or noncommutative) at the Planck scale. 
The main aim of this talk is 
to show that a short-distance cutoff or noncommutativity 
can suppress the CMB anisotropies at large angular scales \cite{fkm1}.

This claim would be against one's intuition, 
because such short-distance structure usually does not 
give rise to important effects on large-scale physics 
in local quantum field theories. 
However, in the exponentially expanding universe as in inflationary models, 
a cutoff on comoving modes can be monotonically increasing function of time 
and, as we see below, 
can suppress large-scale modes when a particular class 
of cutoff is chosen.

Before proceeding to a discussion on this ``large-scale damping," 
we here briefly review the basic ingredients 
in inflationary models.

The flat FRW metric is 
 $ds^2=-dt^2+a^2(t)\,d{{\bm x}}^2$.
By making a Fourier-transform of an inflaton field as 
$\phi(t,{\bm x})=\sum_{\bm k}e^{i{\bm k}\cdot{\bm x}}\,\phi_{\bm k}(t)$, 
the physical wave length is given by multiplying 
the comoving wave length $1/k$ by the scale factor, 
and thus increases monotonically in the expanding universe:
\begin{align}
 \lambda_{\rm phys}(t)=\frac{a(t)}{k}. 
\end{align} 
On the other hand, the Hubble parameter is defined to be 
 $H(t)\equiv \dot{a}(t)/a(t)$.  
During inflation 
$\big( a(t)=(1/H)\,e^{H(t-t_i)}$ 
($t_i$: initial time of inflation)$\big)$, 
the Hubble length $1/H$ is constant, 
and thus for each mode $\phi_{\bm k}(t)$ 
there is the moment $t^{\rm C}_{\bm k}$ 
at which $\lambda_{\rm phys}(t^{\rm C}_{\bm k})= 1/H$.
Since the Hubble length gives the physical (proper) distance 
scale beyond which two points cannot be causally correlated,  
the mode $\phi_{\bm k}(t)$ becomes classical 
after crossing the Hubble horizon, $\lambda_{\rm phys}(t)\gtrsim 1/H$.
By using the cosmological perturbation theory \cite{Liddle}, 
the angular power spectrum of the CMB observed at the present time 
can be shown, at large angular scales, to be proportional to 
that of inflaton evaluated at the exit of inflation. 
In this sense, what we observe as the CMB anisotropies 
are the classical ``fossils" of the quantum fluctuations of inflaton. 
Note that the smaller the comoving wave number $k$ is, 
the earlier the corresponding mode crosses the horizon,  
and thus, the large-scale modes (with small $k$) 
become classical at early times during inflation. 


\section{\large{A mechanism of the large-scale damping}}

Now we explain how a short-distance cutoff 
can affect the large-scale behavior \cite{fkm2}. 

\begin{figure}[ht]
\begin{center}
  \includegraphics[width=12cm]{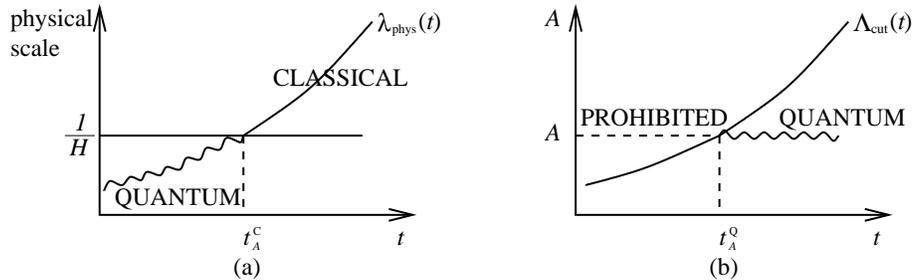}
\end{center}
\vspace{-1cm}
\caption{\footnotesize{(a) $t^{\rm C}_A$ is the moment 
at which a mode $A$ crosses the horizon and becomes classical. 
(b) $t^{\rm Q}_A$ is the moment when the mode $A$ 
starts its quantum fluctuation.}}
\label{fig1}
\end{figure}

In order to give a general discussion, 
we consider a mode expansion of the inflaton field $\phi(t,{\bm x})$ 
in a generic form: 
\begin{align}
 \phi(t,{\bm x})=\sum_{A}\,\big( a_A\,\psi_A(t,{\bm x})+
  a_A^\dagger\,\psi_A^*(t,{\bm x})\big), 
\end{align}
where $\{A\}$ is a set of comoving modes, 
and $\psi_A(t,{\bm x})$'s are symplectically orthonormal, 
positive-energy solutions 
to the Klein-Gordon equation on the curved spacetime, 
 $ds^2=-dt^2+(1/H^2)\,e^{2H(t-t_i)}\,d{\bm x}^2$.
This field is properly quantized by setting canonical commutation 
relations as $\big[a_A,a_B^\dagger\big]=\delta_{AB}$. 
We introduce a pseudo-order in the set $\{A\}$ 
such that small $A$ implies a larger scale. 
An example is to simply set $A={\bm k}$ and 
require that $A={\bm k}<A'={\bm k'}$ when $|{\bm k}|<|{\bm k'}|$.
A mode $A$ starts its quantum fluctuation at an early time, 
and becomes classical around the moment of crossing the horizon,  
which we set $t=t^{\rm C}_A$ (see Fig.~\ref{fig1}~(a)). 

On the other hand, in the presence of short-distance cutoff, 
the comoving modes have a cutoff on large $A$, 
$A\leq \Lambda_{\rm cut}$. 
As we see below, this cutoff $\Lambda_{\rm cut}$ 
can be a monotonically increasing function in time 
in the expanding universe, $\Lambda_{\rm cut}=\Lambda_{\rm cut}(t)$. 
Thus, to a given mode $A$ 
one can assign the moment $t^{\rm Q}_A$ at which the mode $A$ 
starts its quantum fluctuation (see Fig.~\ref{fig1}~(b)). 
Before the moment ($t<t^{\rm Q}_A$) the mode $A$ is prohibited 
to exist as a quantum fluctuation.

Suppose that there exists the following inequality with some mode $A_c$: 
\begin{align}
 A \lessgtr A_c ~\Longleftrightarrow~ t^{\rm C}_A \lessgtr t^{\rm Q}_A. 
 \label{inequality}
\end{align}
Then the modes $A<A_c$ and the modes $A>A_c$ behave quite differently: 

\noindent(1) $A<A_c$ (larger-distance scale):
In this case, the mode $A$ must become classical 
before its quantum fluctuation starts, 
and thus the classical amplitude must be largely suppressed, 
because no fossil can exist if the mode does not have a life of 
quantum fluctuation.  
We thus expect that the corresponding temperature fluctuation 
in the CMB has large suppression. 

\noindent(2) $A>A_c$ (shorter-distance scale):
In this case, the mode $A$ has a period of quantum fluctuation, 
leaving a classical fossil in finite size after crossing the horizon.  
Since the effect of cutoff is expected to disappear rapidly, 
the classical value will be almost the same 
with that in the case without cutoff. 
We thus expect that the corresponding temperature fluctuation 
has almost the same magnitude with that in the absence of cutoff.

Thus, with a cutoff satisfying the condition (\ref{inequality}), 
the mean square of the classical amplitude will have 
a sharp damping on the large-scale modes $A<A_c$. 
(Actually, the process for a mode to become classical 
proceeds only gradually around the moment $t^{\rm C}_A$, 
so that the damping will be smeared \cite{fkm2}.)


\section{\large{Example -- fuzzy sphere}}

We give a model of noncommutative inflation 
satisfying the condition (\ref{inequality}). 
We here introduce the noncommutativity only to the angular coordinates 
\cite{fkm1} 
because their noncommutativity will be most relevant 
to the large-scale damping in the angular power spectrum of the CMB. 
In fact, in the cosmological perturbation theory, 
one can consider the time evolution of each mode separately. 
{}Furthermore, the angular power spectrum of the CMB anisotropy 
can be related to the fluctuations of gravitational potential 
on the last scattering surface, 
and thus is sensitive to the fluctuations only in the angular directions. 
Since the noncommutativity to a given direction 
is expected to give its major effects to the fluctuations 
in the corresponding direction, 
the introduction of noncommutativity to the other directions 
(time and radius) 
will not give a drastic change to the angular power spectrum.%
\footnote{
Recently Tsujikawa, Maartens and Brandenberger \cite{Tsujikawa:2003gh} 
and Huang and Li \cite{Huang:2003hw} 
have analyzed the noncommutative inflation 
introducing the noncommutativity only to time and radial coordinates, 
and shown that the effect is not strong enough 
to give a sharp damping at large angular scales. 
}

{}For later convenience, 
we introduce the conformal time $\eta$, 
\begin{align}
 ds^2=a^2(\eta)\big(-d\eta^2+dr^2+r^2d\Omega^2\big), 
\end{align}
for which the scale factor is expressed as 
\begin{align}
 a(\eta)=-\frac{1}{H\eta}\quad(\eta<0). 
\end{align}
Note that $\eta$ is negative during inflation, 
and the exit time of inflation is given by 
taking $\eta\rightarrow -0$.

We introduce a fuzzy sphere 
such that there are at most one-bit degrees of freedom 
in the physical area $L_{\rm cut}^2$. 
Since the physical area of the sphere for each $(\eta,r)$ 
is $4\pi\big(a(\eta)r\big)^2$, 
the maximum degrees on the fuzzy sphere is given by \cite{fkm1} 
\begin{align}
 \frac{4\pi\big(a(\eta)r\big)^2}{L_{\rm cut}^2}
  =\frac{4\pi r^2}{(L_{\rm cut}H)^2\eta^2}
  \equiv N(\eta,r)+1,  
\end{align}
which is actually a monotonically increasing function of time $\eta$ 
(recall that $\eta<0$). 
Since the mode expansion of a scalar field on the fuzzy sphere 
has a limiting multipole, $l\leq \Lambda_{\rm cut}=N(\eta,r)$ 
\cite{fkm1}, 
the inflaton field is expanded as 
\begin{align}
 &\phi(\eta,r,\Omega)=\sum_{l=0}^{N(\eta,r)}\sum_{m=-l}^{l}\,
  \int_0^\infty \frac{dk}{2\pi}\,
  \big( a_{klm}\,\psi_{klm}(\eta,r,\Omega)+{\rm h.c.}\big),\\ 
 &\psi_{klm}(\eta,r,\Omega)=H\sqrt{\frac{2}{k}}\,\big(1+ik\eta\big)\,
  e^{-ik\eta}\,j_l(kr)\,Y_{lm}(\Omega). 
\end{align}
The moment $\eta^{\rm C}_A$ at which the mode $A=(k,l,m)$ 
becomes classical is obtained 
by setting $a(\eta^{\rm C}_A)/k\equiv 1/H$: 
\begin{align}
 \eta^{\rm C}_A=-\frac{1}{k}, 
\end{align}
while the moment $\eta^{\rm Q}_A$ at which the mode 
starts its quantum fluctuation is given 
by setting $l\equiv N(\eta^{\rm Q}_A,r)$:
\begin{align}
 \eta^{\rm Q}_A=-\frac{1}{L_{\rm cut}H}\,\sqrt{\frac{4\pi}{l+1}}\,r. 
\end{align}
The critical mode $A_c$ can be roughly estimated 
by noting that $\eta^{\rm C}_A$ can be approximated 
by $-2r/\pi(l+1)$ 
because the spherical Bessel function $j_l(kr)$ 
has a sharp peak at $kr=(l+1)\pi/2$. 
Thus, setting $\eta^{\rm Q}_{A_c}=\eta^{\rm C}_{A_c}$, 
we have 
\begin{align}
 l_c=\frac{(L_{\rm cut}H)^2}{\pi^3}-1. 
\end{align}
It is easy to see that the condition (\ref{inequality}) holds 
with this $l_c$, 
and thus we expect that the angular power spectrum will have 
a sharp damping for $l\lesssim l_c$. 

\begin{figure}[ht]
 \begin{center}
  \rotatebox{270}{
  \resizebox{!}{75mm}{\includegraphics{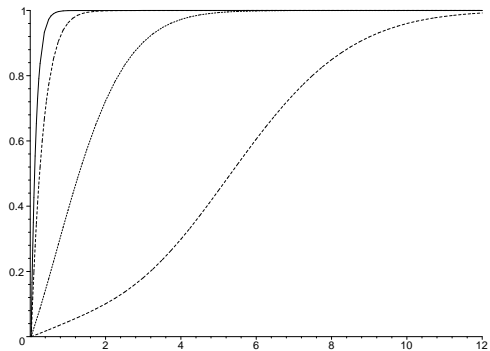}}}
 \end{center}
\vspace{-1cm}
\caption{\footnotesize{The resulting damping factor as a function 
of multipole $l$. 
Here the spectral index is set to be $n=0.95$. 
$L_{\rm cut}H$ is 0.1, 1, 5, 10 from top to bottom \cite{fkm1}.}} 
\label{N=0.95}
\end{figure}

A more detailed analysis along the above argument 
is made in Refs.\ \cite{fkm1} and \cite{fkm2} 
with a constant spectral index $n$. 
{}For $n=0.95$, the resulting damping factor 
is given in Fig.~\ref{N=0.95}.

\section{\large{Conclusion and outlook}}

We have shown that a class of short-distance cutoff 
(or noncommutativity) can give rise to 
suppression on large-scale modes. 
In order to have a damping for $l\lesssim 10$, 
the noncommutative scale $L_{\rm cut}$ 
is of the same order to the Hubble scale $1/H$. 
In Ref.\ \cite{fkm2}, the model using a fuzzy sphere is further 
investigated, by setting the initial condition on the inflaton fluctuations 
such that every mode starts its quantum fluctuation 
as the vacuum fluctuation with respect to the Hamiltonian 
upon being released from the constraint of cutoff. 
We also sketch a holographic rationale 
for the cutoff using a fuzzy sphere.  
An investigation along this direction is now in progress 
and will be reported elsewhere.


\section*{\large{Acknowledgments}}

M.F.\ would like to thank the organizers of QTS3. 
This work is supported in part by Grant-in-Aid (No.\ 15540269) 
from the Japan Ministry of Education, Culture, Sports, 
Science and Technology, 
and by Grant-in-Aid for the 21st Century COE 
``Center for Diversity and Universality in Physics."


%
%
%
%
%
%


\baselineskip=0.8\normalbaselineskip


\begin{thebibliography}{99}


\bibitem{data}
J.~R.~Bond, A.~H.~Jaffe and L.~Knox,
{\it Phys.\ Rev.\ D} {\bf 57}, 2117 (1998) 
[arXiv:astro-ph/9708203]; 
C.~L.~Bennett {\it et al.},
arXiv:astro-ph/0302207; 
D.~N.~Spergel {\it et al.},
arXiv:astro-ph/0302209;
H.~V.~Peiris {\it et al.},
arXiv:astro-ph/0302225.


\bibitem{Liddle}
A.~R.~Liddle and D.~H.~Lyth,
{\it ``Cosmological Inflation and Large-Scale Structure,''}
Cambridge University Press (2000).


\bibitem{minimum_length}
G.~Veneziano,
{\it Europhys.\ Lett.} {\bf 2}, 199 (1986);
D.~J.~Gross and P.~F.~Mende,
{\it Nucl.\ Phys.\ B} {\bf 303}, 407 (1988);
D.~Amati, M.~Ciafaloni and G.~Veneziano,
{\it Phys.\ Lett.\ B} {\bf 216}, 41 (1989);
R.~Guida, K.~Konishi and P.~Provero,
{\it Mod.\ Phys.\ Lett.\ A} {\bf 6}, 1487 (1991);
T.~Yoneya, in {\it ``Wandering in the Fields,''} 
eds. K.~Kawarabayashi, A.~Ukawa (World Scientific, 1987 ), p.419;
M.~Li and T.~Yoneya,
{\it Phys.\ Rev.\ Lett.}  {\bf 78}, 1219 (1997) 
[arXiv:hep-th/9611072];
T.~Yoneya,
{\it Prog.\ Theor.\ Phys.}  {\bf 103}, 1081 (2000)
[arXiv:hep-th/0004074].


\bibitem{fkm1}
M.~Fukuma, Y.~Kono and A.~Miwa, 
to appear in {\it Nucl.\ Phys.\ B} 
[arXiv:hep-th/0307029]. 


\bibitem{fkm2}
M.~Fukuma, Y.~Kono and A.~Miwa,
arXiv:hep-th/0312298.


\bibitem{Tsujikawa:2003gh}
S.~Tsujikawa, R.~Maartens and R.~Brandenberger,
{\it Phys.\ Lett.\ B} {\bf 574}, 141 (2003) 
[arXiv:astro-ph/0308169].


\bibitem{Huang:2003hw}
Q.~G.~Huang and M.~Li,
{\it JCAP} {\bf 0311}, 001 (2003) 
[arXiv:astro-ph/0308458].


\end{thebibliography}
\end{document}